\pdfoutput=1
\documentclass[
reprint,
superscriptaddress,
nofootinbib,
amsmath,amssymb,
aps,
prd,
floatfix,
]{revtex4-2}

\usepackage[table,svgnames,dvipsnames]{xcolor}
\definecolor{Blue}{HTML}{1c1e94}
\usepackage[normalem]{ulem}
\usepackage{aas_macros}
\usepackage{graphicx}
\usepackage{dcolumn}
\usepackage{bm}
\usepackage[mathlines]{lineno}
\usepackage{orcidlink}
\usepackage{cases}
\usepackage{booktabs}
\usepackage{comment}
\usepackage{multirow}
\usepackage{makecell}
\usepackage{siunitx}
\usepackage{booktabs}
\graphicspath{{figs/}}
\usepackage{fontawesome5}
\usepackage{xspace}
\usepackage{hyperref}

\def\be{\begin{equation}}
\def\ee{\end{equation}}

\def\ba#1\ea{\begin{align}#1\end{align}}

\newcommand{\code}[1]{{\texttt{#1}}}
\renewcommand{\emph}[1]{\textit{#1}}

\def\nefertiti{\code{Nefertiti}\xspace}

\usepackage[nameinlink,noabbrev]{cleveref}
\crefname{equation}{Eq.}{Eqs.}
\crefname{section}{Section}{Sections}
\crefname{figure}{Fig.}{Figs.}
\crefname{table}{Table}{Tables}
\crefname{appendix}{Appendix}{Appendices}
\Crefname{figure}{Figure}{Figures}
\Crefname{equation}{Equation}{Equations}
\Crefname{section}{Section}{Sections}
\Crefname{table}{Table}{Tables}
\newcommand{\refeq}[1]{Eq.~(\ref{eq:#1})}
\newcommand{\refeqs}[2]{Eqs.~(\ref{eq:#1})--(\ref{eq:#2})}
\newcommand{\reffig}[1]{Fig.~\ref{fig:#1}}
\newcommand{\reffigs}[2]{Figs.~\ref{fig:#1}--\ref{fig:#2}}

\newcommand{\refsec}[1]{Sec.~\ref{sec:#1}}

\newcommand{\press}{\frac{p}{c^2}}
\newcommand{\cssqnorm}{\frac{c_s^2}{c^2}}

\newcommand{\Mpch}{\,h^{-1}\mathrm{Mpc}}

\newcommand{\hmpcinv}{\,h\,{\mathrm{Mpc}^{-1}}}

\newcommand{\vel}{\bm{v}}

\newcommand{\der}[2]{\frac{\partial #1}{\partial #2}}

\newcommand{\Om}{\ensuremath{\Omega_{{\rm m}}}\xspace}

\begin{document}

\title{Cosmological hydrodynamical simulations of clustering dark energy with Nefertiti}

\author{Linda Blot \orcidlink{0000-0002-9622-7167}} \email{linda.blot@ipmu.jp}
\affiliation{Center for Data-Driven Discovery, Kavli IPMU (WPI), UTIAS, The University of Tokyo, Kashiwa, Chiba 277-8583, Japan}
\affiliation{Laboratoire d’etude de l’Univers et des phenomenes eXtremes, Observatoire de Paris, Université PSL, Sorbonne Université, CNRS, 92190 Meudon, France}
\author{Théo Gayoux \orcidlink{0009-0008-9527-1490}}
\affiliation{Laboratoire d’etude de l’Univers et des phenomenes eXtremes, Observatoire de Paris, Université PSL, Sorbonne Université, CNRS, 92190 Meudon, France}
\affiliation{Universit\'e Paris-Cit\'e, 5 Rue Thomas Mann, 75013, Paris, France}
\author{Fabian Schmidt\orcidlink{0000-0002-6807-7464}} \email{fabians@mpa-garching.mpg.de}
\affiliation{Max–Planck–Institut für Astrophysik, Karl–Schwarzschild–Straße 1, 85748 Garching, Germany}
\author{Pier Stefano Corasaniti \orcidlink{}}
\affiliation{Laboratoire d’etude de l’Univers et des phenomenes eXtremes, Observatoire de Paris, Université PSL, Sorbonne Université, CNRS, 92190 Meudon, France}
\author{Bastien de Ligondes}
\affiliation{École Polytechnique, Rte de Saclay, 91120 Palaiseau, France}

\date{\today}

\begin{abstract}
We present the first cosmological simulations that consistently include nonlinear clustering dark energy evolved as a fluid with the numerical hydrodynamics code \code{Nefertiti}. Dark energy perturbations become fully nonlinear on small scales, developing significant density fluctuations without exhibiting the catastrophic instabilities previously reported.
We show results for the density distribution, power spectrum, and halo profiles of dark energy. Clustering dark energy contributes to the total density perturbation at the $\sim 10\%$ level inside and around massive halos in our simulations with constant $w=-0.9$, a significant potential signal for lensing and dynamical probes. These simulations pave the way to robust constraints on the speed of sound of dark energy perturbations from large-scale structure data.
\end{abstract}

\maketitle

\textit{Introduction.}---The nature of dark energy remains entirely unknown, and one of the most significant open problems in physics. If dark energy is anything other than a cosmological constant, it must possess stress-energy perturbations, which interact gravitationally with perturbations in the dark and ordinary matter. An important class of dark energy models exhibits a small speed of sound $c_s \ll c$---hereafter referred to as \emph{clustering dark energy}---which implies that dark energy perturbations are not restricted to very large (horizon) scales, but indeed grow on all cosmological scales. This class of models includes k-essence \cite{chiba/okabe/yamaguchi,ArmendarizPicon:2000ah}, a scalar field theory with a non-canonical kinetic term. Moreover, any stable theory of dark energy that allows for crossing the phantom divide, where $w\equiv\bar p/\bar \rho$ becomes less than $-1$, \emph{must} have extremely small speed of sound \cite{Creminelli:2006xe, Creminelli:2008wc}. 
The recent results from the DESI collaboration, when combined with Supernova Ia data, show a preference for dark energy models characterised by a phantom divide crossing \cite{DESI-DR1-BAO,DESI-DR1-FS,DESI-DR2-BAO}. If these results are confirmed
in the future,
they will provide additional motivation for studying dark energy with a small speed of sound.

Clustering dark energy can leave interesting signatures in the large-scale structure (LSS) on both large and small scales, including galaxy clustering, dynamical probes, and gravitational lensing. The small-scale regime is especially interesting, as this is generally where LSS observations are most precise. However, dark energy perturbations are expected to become non-linear on small scales \cite{Creminelli:2009mu,Sefusatti:2011JCAP}. Even though their amplitude is not as high as that of perturbations in cold and ordinary matter (see below), a consistent solution for the joint evolution of both clustering dark energy and matter, coupled by gravity, will require full simulations.

The numerical techniques for solving for non-linear dark energy perturbations and those of the dark matter are fundamentally different, requiring the development of dedicated codes. To avoid having to develop a dedicated code for each dark energy model, two different paradigms for an \emph{effective} treatment of dark energy perturbations can be followed: one is based on the \emph{effective field theory (EFT)} of dark energy \cite{Gubitosi:2012hu}, which identifies a scalar degree of freedom $\pi$ as the Goldstone boson of time translations. 
A wide range of dark energy models is described by two effective functions of time: the quantity $w(\tau)$ and a function $M_2(\tau)$, which together describe the background pressure as well as pressure perturbations. 
This formulation has been adopted in \cite{hassani$k$evolutionRelativisticNbody2019} and implemented in the \code{k-evolution} code.
The resulting field equation for $\pi$ is a quasilinear hyperbolic PDE that can be solved numerically on a grid.
In this formulation, there are non-linear terms $\propto (\boldsymbol{\nabla}\pi)^2$ that are enhanced in the small-speed-of-sound limit, and which have been identified as the source of instabilities in the solution for $\pi$ \cite{2022PhRvD.105b1304H,hassaniGenericInstabilityClustering2023}.

A fundamental alternative is to treat the dark energy as an \emph{effective fluid}, again described by two functions of time, $w(\tau)$ and $c_s(\tau)$. 
In the effective fluid approach, $c_s^2 = \partial \delta p/\partial\delta\rho$ strictly describes the relation between pressure perturbations and density perturbations.  
The limit $c_s\to 0$ thus is in one-to-one correspondence with vanishing pressure gradients, which implies a dark energy fluid that co-moves with dark matter on large scales. As originally proposed in \citet{Blot:2022mnt}, the effective fluid approach enables the use of advanced techniques from computational fluid dynamics, in particular finite-volume methods that work via conserved fluxes and can deal with strong gradients (shocks and contact discontinuities). This is the approach taken in \nefertiti, a fork of the \code{Ramses} code \cite{teyssier2002} that implements a fluid description of clustering dark energy, and that was used to perform the cosmological simulations presented here.

In our simulations, we do not encounter instabilities of the sort found in \cite{2022PhRvD.105b1304H,hassaniGenericInstabilityClustering2023}; that is, the solution for the dark energy fluid remains well-behaved over the duration of the simulation. We will discuss possible reasons for these differences in the conclusions.

It is also worth mentioning approaches that treat dark energy perturbations at linear order, by including a linear-order Boltzmann-code prediction as a source term in standard matter-only cosmological simulations. This has been implemented in the \code{CONCEPT} and \code{PKDGRAV3} codes \citep{2019JCAP...08..013D}. Such approaches break down when dark energy perturbations become non-linear, which is generally the case for clustering dark energy.

\begin{figure}[!t]
   \includegraphics[width=1.0\linewidth]{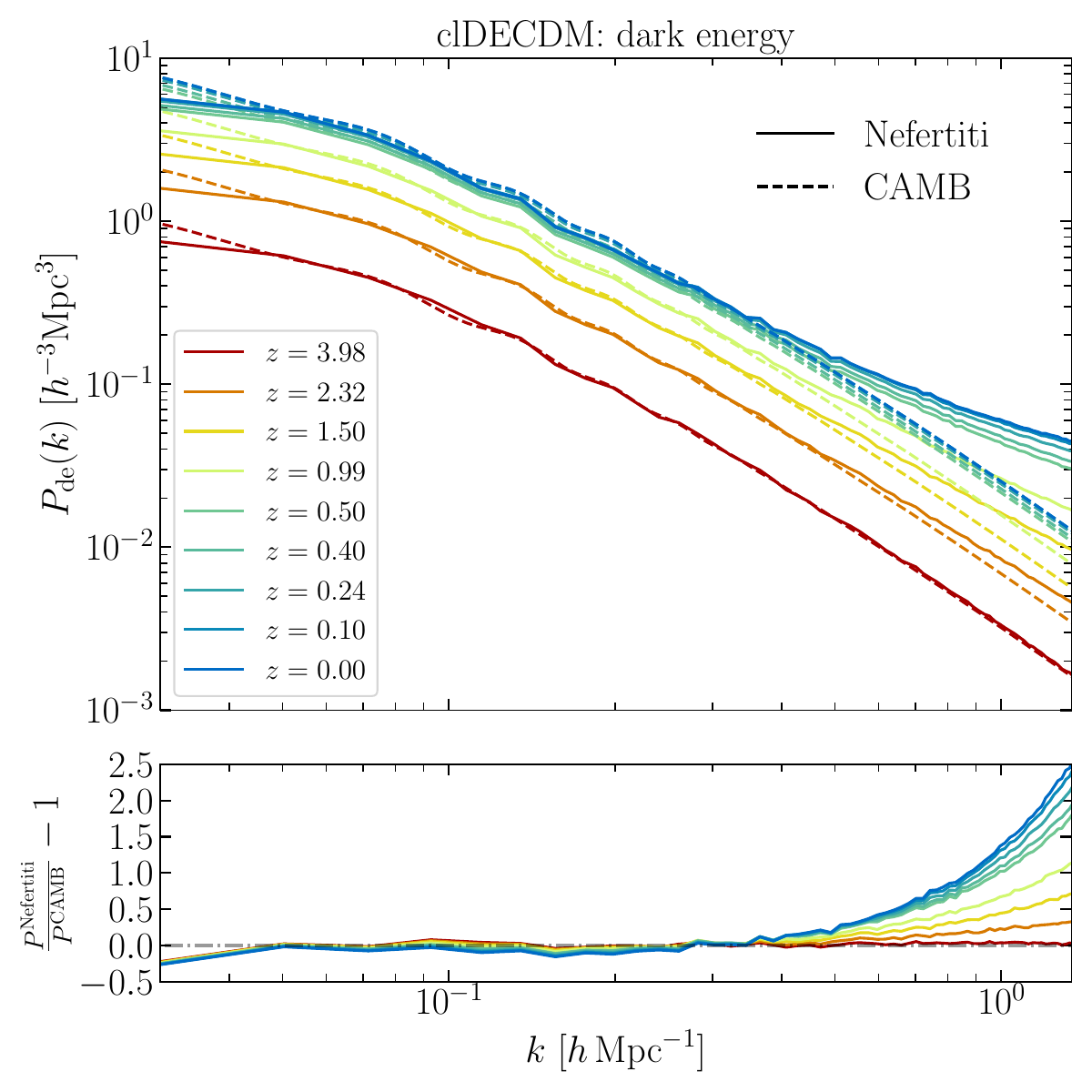}
   \caption{Top panel: Non-linear dark energy power spectrum between redshift 4 and 0 from \nefertiti (continuous lines), and linear prediction from \code{CAMB} (dashed lines). Bottom panel: relative difference between the nonlinear and linear power spectra.}
   \label{fig:pk_de}
 \end{figure}
 
 \begin{figure}[!t]
   \includegraphics[width=1.0\linewidth]{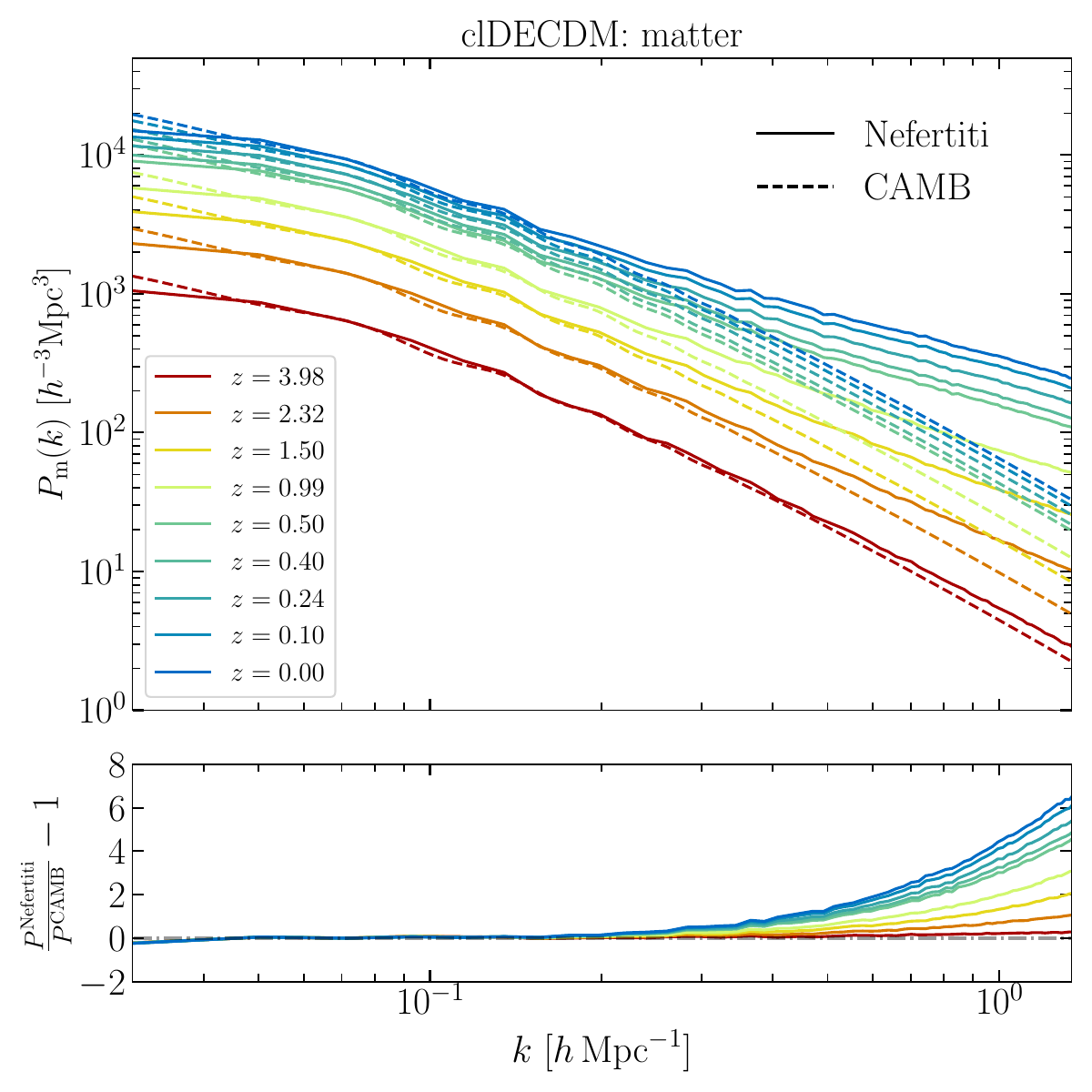}
   \caption{Top panel: Non-linear matter power spectrum between redshift 4 and 0 from \nefertiti (continuous lines) and linear predictoin from \code{CAMB} (dashed lines). Bottom panel: relative difference between the nonlinear and linear power spectra.}
   \label{fig:pk_dm}
 \end{figure}
 
\medskip

\textit{Dark energy fluid description.}---In this work, we choose to represent dark energy as a fluid with an equation of state (EoS):
\be
p= w\, \bar{\rho}\, c^2 + c_s^2\, \delta\rho\,,
\label{eq:eos}
\ee
which relates the total dark energy pressure $p$ to the mean dark energy density $\bar{\rho}$ and the density perturbation $\delta\rho=\rho-\bar{\rho}$ using two parameters: $w$, which dictates the behaviour of the background, and $c_s$, which regulates the clustering on small scales. In the limit of small speed of sound, matter and dark energy are comoving, so we can ignore additional terms coming from frame transformations.
\refeq{eos} contains the leading terms in an expansion in perturbations and spatial derivatives. Higher spatial derivatives are associated with the mass scale of the underlying microphysical model of dark energy, and are thus expected to be irrelevant on cosmological scales. Corrections to the equation of state involving $(\delta\rho)^2$ could be present, but we assume that they are parametrically suppressed with the same $c_s^2$, so that they likewise become irrelevant as $c_s^2 \to 0$. This has been shown to be the case for the ghost condensate, for example \cite{Arkani-Hamed:2003pdi,Arkani-Hamed:2005teg},
a particular realisation of a clustering dark energy scenario, where
the full equation of state is given as $p(\rho) = \rho^2/(2M_{\rm g.c.}^4)$.
In this case, the next term in the expansion \refeq{eos} is given by $c_s^2 (\delta\rho)^2/\bar\rho$ which can be neglected in the $c_s \to 0$ limit.

In an expanding background, the evolution of the fluid is described by the continuity and Euler equations
\ba
&\der{\rho}{\tau} + 3\mathcal{H}\left(\rho + \press \right) + \boldsymbol{\nabla} \cdot \left(\rho + \press \right) \vel = 0\,,\label{eq:cont}\\
&\der{v}{\tau} + \mathcal{H}\vel + \left( \vel \cdot  \boldsymbol{\nabla} \right)  \vel + \boldsymbol{\nabla} \phi = - \frac{\boldsymbol{\nabla} p}{\rho + \press}\,,\label{eq:eul}
\ea
where $\rho$ is the total dark energy density, $\tau$ is the conformal time, $\mathcal{H}$ is the Hubble parameter, $\vel$ is the peculiar velocity, and $\phi$ is the gravitational potential. Dark energy is coupled to matter only through gravitational interactions, which are accounted for in the Poisson equation
\be
\nabla^2\phi=4\pi\, G \,a^2\left( \delta\rho_m +\delta\rho + 3 \frac{\delta p}{c^2}  \right)\,, \label{eq:possion}
\ee
where $\delta\rho_m$ is the overdensity of matter.
From \refeqs{cont}{eul} we can derive a set of equations in quasi-conservative form that can be more easily implemented in the \nefertiti code. The actual form of the equations that are solved numerically can be found in \refsec{equations} of the supplementary material. Note that since dark energy is comoving with matter, its typical velocity is that of large-scale-structure flows. Hence, we are here exploring the regime where the speed of sound is very small compared to the typical fluid velocity, and we expect the fluid to develop large-scale coherent shocks, especially in the vicinity of collapsed structures.

\medskip

\textit{Nefertiti code.}---To solve the system of equations \refeqs{eos}{possion}, we implement the numerical methods presented in~\citet{Blot:2022mnt} in \nefertiti, coupled with a source splitting method to time-integrate the source terms (see \refsec{sources} of the supplementary material for details). We use finite-volume hydrodynamical solvers, since they are known to provide more accurate results in terms of conservation of physical quantities, and in the presence of discontinuities such as shocks or rarefaction waves. Fluxes are computed with the MUSCL-Hancock method \cite{VanLeer1984}, using the Linear Primitive Variable Riemann Solver (LPVRS) \cite{Blot:2022mnt}.
The code inherits the Adaptive Mesh Refinement implementation of \code{Ramses}, where the density and gravitational potential are interpolated on a multi-level grid, and the Poisson equation is solved via relaxation methods. Following the approach originally adopted in \cite{2010MNRAS.401..775A}, we implemented the possibility of reading the tabulated time evolution of background quantities from an external file, which are then interpolated at the necessary time internally in \nefertiti.

  \begin{figure}[!t]
   \includegraphics[width=1.0\linewidth,trim=0cm 0cm 1cm 1cm,clip=true]{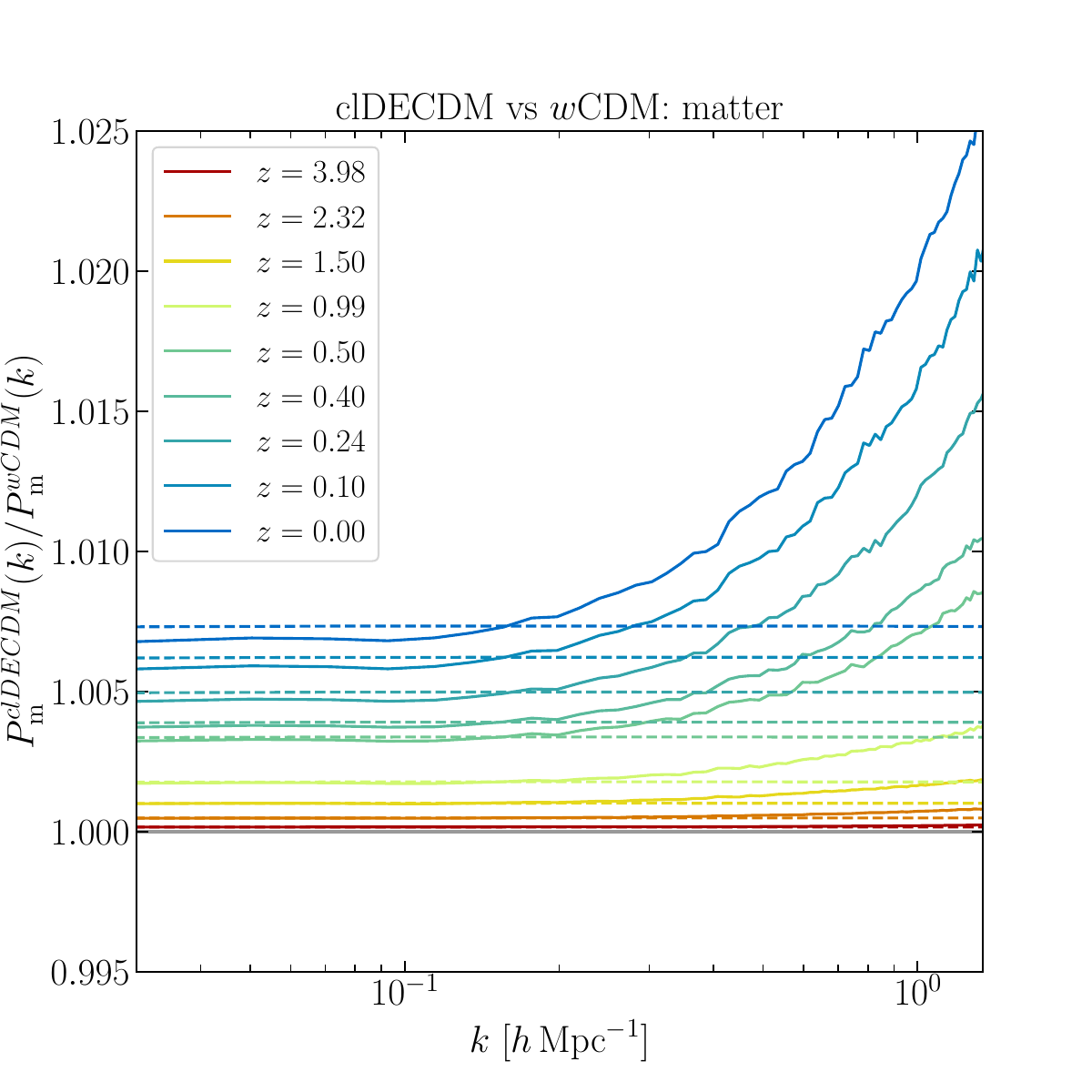}
   \caption{Ratio of the matter power spectrum in clustering dark energy (clDECDM) and $w$CDM cosmologies from \nefertiti (continuous lines) and \code{CAMB} (dashed lines).}
   \label{fig:pk_dm_cosmo_ratio}
 \end{figure}

\medskip

\textit{Simulations.}---To allow for direct comparison with the \code{k-evolution} results of \citet{hassaniGenericInstabilityClustering2023}, we run simulations with a box-size of $300 \Mpch$, a fixed $512^3$ grid (i.e. without grid refinement), following the evolution of $512^3$ matter particles. This setup yields a mass resolution of $m_{\rm p} \simeq 1.7\times10^{10} \,M_\odot\,h^{-1}$ and a spatial resolution of $\Delta x \simeq 0.6\,\rm Mpc\,h^{-1}$. We ran a $w$CDM simulation, where the effect of dark energy is only present at the level of background evolution, in addition to the clustering dark energy (clDECDM) simulation, where dark energy perturbations are also evolved as a fluid on the grid. Both simulations employ the same time-stepping criterion, $\Delta \log{a}=5\times 10^{-3}$, to allow an accurate comparison between the two cosmologies. This value was chosen based on convergence tests for the clustering dark energy simulation. Both simulations share the same values of the cosmological parameters: $\Om=0.307110$, $h=0.6777$, $w=-0.9$, with the exception of the dark energy speed of sound, which is $c_s^2=10^{-9}\,c^2$ for clDECDM and, essentially, $c_s^2=c^2$ for $w$CDM. We generate the initial conditions at $z=40$ using the \code{grafic} code \citep{2008ApJS..178..179P}, which we have modified to obtain initial conditions for the dark energy fluid using the linear relation:
\be
\delta_{\rm de}=\frac{1+w}{1-3w}\delta_{\rm m},
\ee
where $\delta=\delta\rho/\bar{\rho}$. This relation is valid in the matter-dominated era of the Universe's expansion and in the $c_s\rightarrow0$ regime; transients induced by this approximation are expected to be at the sub-percent level. We leave a more accurate treatment of the initial conditions, including the scale dependence of the growth factor in case of finite speed of sound, for future work. 
We identify halos in the particle field using the \texttt{rockstar} \cite{Behroozi_2012} halo finder. The \code{k-evolution} simulation with the same cosmological parameters, box size, and resolution breaks at $z=8$ \citep{hassaniGenericInstabilityClustering2023}. Our clDECDM simulation using \nefertiti does not show this issue and was run until $z=0$.

\begin{figure}[!t]
    \centering
\includegraphics[width=1\linewidth]{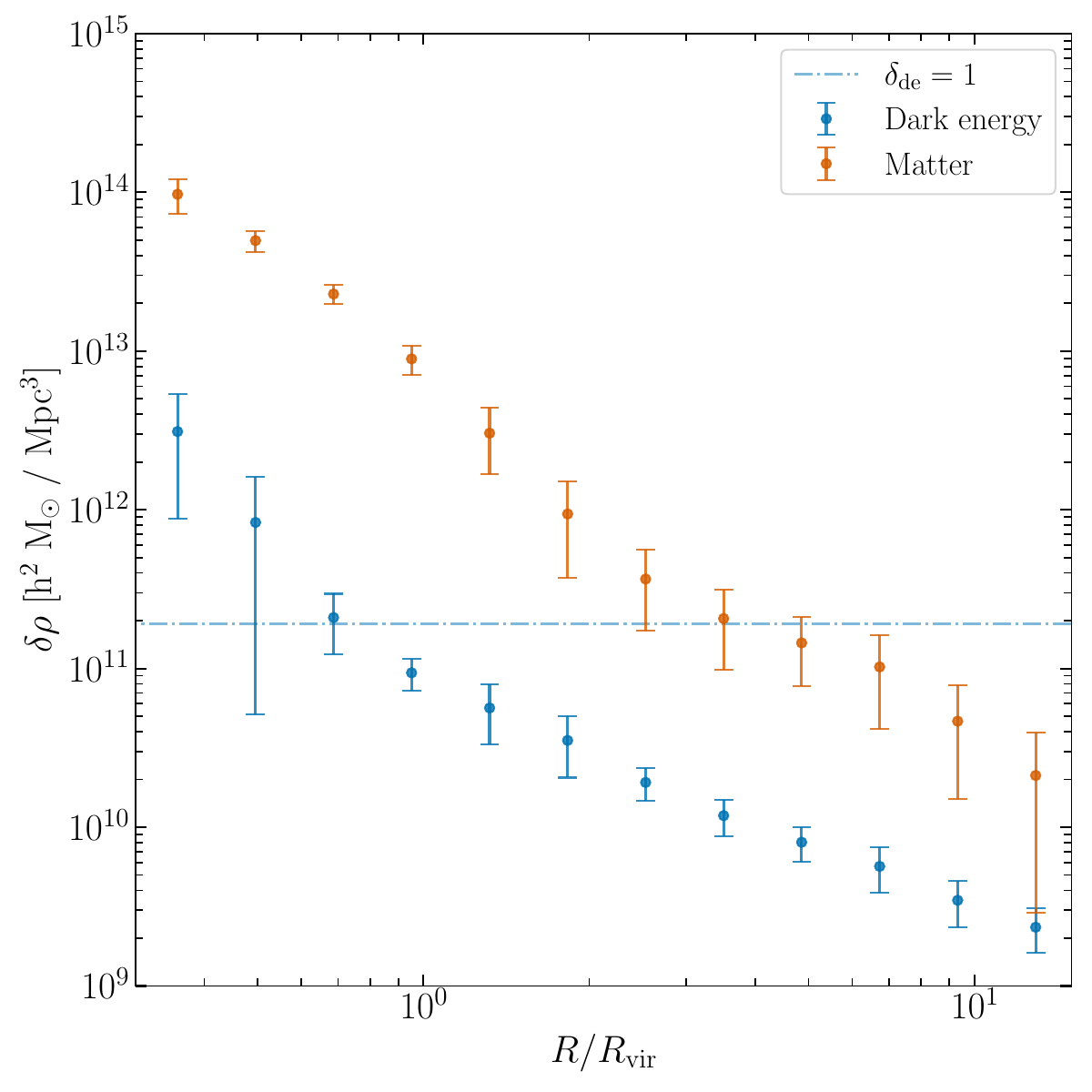}
    \caption{Mean density perturbation profiles for dark energy (in blue) and matter (in orange) as a function of radius, normalised to the virial radius, for the $N_h = 24$ most massive halos in the clDECDM simulation at $z=0$. Error bars are given by the standard deviation across the halo sample. The dash-dotted horizontal blue line shows the density threshold at which $\delta_{\rm de} = 1$. Above this value, dark energy is fully non-linear.}
    \label{fig:profiles}
\end{figure}

\medskip

\textit{Results.}---We show the power spectrum of dark energy and matter for the clDECDM simulation in \reffigs{pk_de}{pk_dm} respectively. In the bottom panels, we plot the relative difference with respect to the linear power spectra computed with \code{CAMB} \citep{Lewis:1999bs}. The large-scale (low-$k$) results show that linear growth is correctly recovered in both matter and dark energy. Both components develop significant non-linearities at small scales that grow towards late times. In particular, for dark energy the relative difference reaches a factor of a few at $k>0.3\hmpcinv$.

Non-linearities also show up in the ratio of the matter power spectrum in the clDECDM and $w$CDM cosmologies shown in \reffig{pk_dm_cosmo_ratio}, where departures from linear predictions are evident for redshifts $z<1$.

\begin{figure*}
    \centering
    \includegraphics[width=1.0\linewidth]{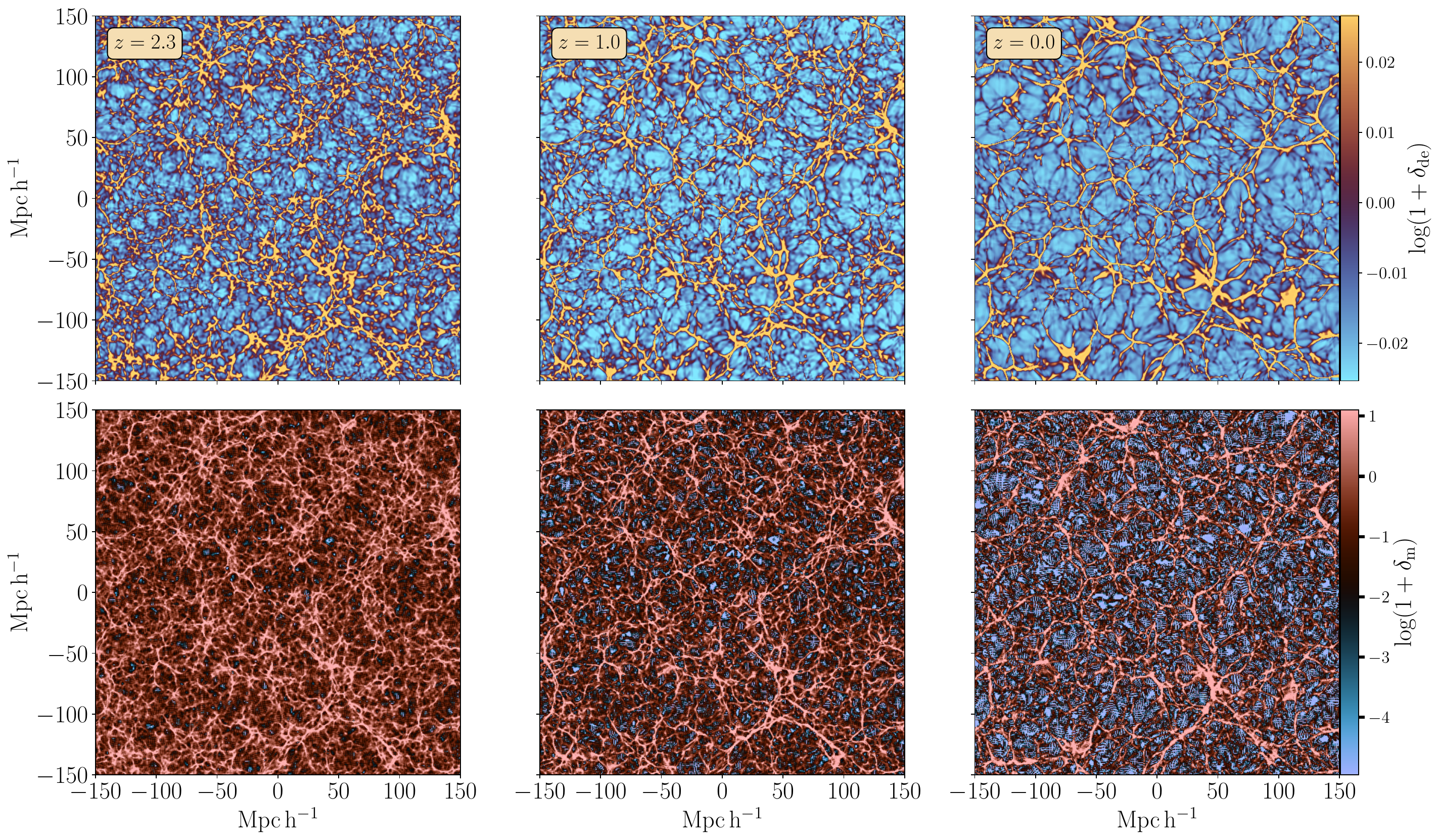}
    \caption{Redshift evolution of DE (upper row) and matter (lower row) overdensity slices predicted by \nefertiti. The redshift corresponding to each snapshot is shown in the top-left corner of each panel of the upper row.}
    \label{fig:density_slices}
\end{figure*}

A visualisation of the large-scale structure in the clDECDM simulation is shown in \reffig{density_slices}, where we can see that the structures in the dark energy density field are highly correlated with those in the matter, as expected.

While the dark energy overdensity remains a few orders of magnitude smaller than the matter one, it does become non-linear in the high-density regions, with values up to $\delta_{\rm de}\approx50$ at $z=0$. This is also shown in Fig.\,\ref{fig:profiles}, which presents the stacked density perturbation profiles for both dark energy (in blue) and matter (in orange) for the $N_h = 24$ most massive halos, for the best-resolved halos with virial masses $M_{\rm vir} > 5\times 10^{14} \, M_\odot\,h^{-1}$. We constructed these profiles by computing the mean density in spherical shells centered on the \texttt{rockstar} halo centers, subtracting the mean cosmic density, and stacking the resulting density perturbation profiles over the $N_h$ halos. The profiles were measured in 12 logarithmically spaced radial bins in the range $[0.3,15]R_{\rm vir}$.
The total overdensity profiles (matter plus dark energy) can be probed by stacked gravitational lensing or dynamical probes. We see that dark energy perturbations contribute at the $\sim 10\%$ level over a broad range of scales.
Moreover, when $R/R_{\rm vir} \lesssim 0.7$, the dark energy density perturbation crosses the non-linear threshold $\delta_{\rm de} = 1$, represented by the dash-dotted blue line. 
On large scales, both overdensity profiles asymptote to zero.

\medskip

\textit{Conclusions.}---We have presented the first simulations of clustering dark energy that use the effective-fluid formulation. Unlike previous work that solved the equation for the scalar Goldstone degree of freedom \cite{2022PhRvD.105b1304H,hassaniGenericInstabilityClustering2023}, we have not found instabilities in our simulations. This could be due to the fact that finite-volume methods can better deal with the strong density/field gradients present in this scenario.
However, it is important to also note that,
due to the non-linear relation between $\delta\rho,\delta p$ and $\pi$ in the EFT of DE (App.~A of \cite{hassani$k$evolutionRelativisticNbody2019}),
$\partial \delta p/\partial\delta\rho$ is not simply described by the parameter $c_{s,\rm EFT}^2$ introduced there. This implies that the $c_s^2\to 0$ limit studied here, and the $c_{s,\rm EFT}^2 \to 0$ limit of \cite{2022PhRvD.105b1304H,hassaniGenericInstabilityClustering2023} might not correspond to the same physical system. We plan to explore a detailed comparison in future work.

The simulations presented here pave the way for using non-linear large-scale structure to constrain clustering dark energy models. This includes galaxy or cluster lensing, redshift-space distortions, and the properties of voids, to name a few possible observables. A crucial next step will be adaptively refining simulations, which will allow for much greater dynamic range in resolution.

\medskip

\textit{Acknowledgments.}---We thank Yann Rasera, Julian Adamek, Farbod Hassani, Martin Kunz, Shinji Mukohyama, and Romain Teyssier for useful discussions. Computations were performed on the HPC system \textit{freya} at the Max Planck Computing and Data Facility. Analysis was performed using the \texttt{yt$\_$astro$\_$analysis} extension
\citep{yt.astro.analysis} of the \texttt{yt} analysis toolkit \citep{yt}, the \texttt{Pylians} \cite{Pylians} Python libraries as well as the \texttt{COLOSSUS} \cite{2018ApJS..239...35D} Python Toolkit.

\bibliographystyle{apsrev4-2}
\bibliography{nefertiti_letter}

\clearpage

\appendix
\onecolumngrid

\widetext
\begin{center}
\textbf{\Large Supplementary Material}
\end{center}

\section{Equations}\label{sec:equations}

In the following, we exclusively deal with dark energy fluid variables, so we continue to denote $\rho \equiv \rho_{\rm de}$, $\vel \equiv \vel_{\rm DE}$. In a generic frame, the fluid equations are \citep{Sefusatti:2011JCAP}
\begin{align}
&\der{\rho}{\tau} + 3\mathcal{H}\left(\rho + \press \right) + \boldsymbol{\nabla} \cdot \left(\rho + \press \right) \vel = 0\label{eq:cont_gf}\,,\\
&\der{\vel}{\tau} + \mathcal{H}\vel + \left( \vel \cdot  \boldsymbol{\nabla} \right)  \vel + \boldsymbol{\nabla} \phi = - \frac{1}{\rho + \press} \left[ \boldsymbol{\nabla} p + \vel \der{}{\tau}\left(\press\right) \right] \label{eq:eul_gf}\,.
\end{align} 

The density perturbation $\delta\rho$ is not gauge invariant and is related to the rest-frame density perturbation $\delta\rho_{\rm rf}$ via
\be
\delta\rho+\dot{\rho}V=\delta\rho_{\rm rf}\,,
\ee
where rf stands for the dark-energy rest-frame, and $V$ is the velocity potential, $\vel=\boldsymbol{\nabla} V$ \citep{BeanDore}. The gauge-invariant relation between pressure and density is
\be
\left(\frac{\delta p}{\dot{p}}-\frac{\delta\rho}{\dot{\rho}} \right)_{\rm rf}=\left(\frac{\delta p}{\dot{p}}-\frac{\delta\rho}{\dot{\rho}} \right)\,,
\ee
from which we can derive, at linear order in pressure perturbations,
\be
\boldsymbol{\nabla} p + \vel \dot{p} = \boldsymbol{\nabla} \left(c_s^2 \delta\rho_{\rm rf}\right)\,.
\ee
This means that in the dark energy rest-frame we can write the fluid equations and the equation of state as 
\begin{align}
&\der{\rho}{\tau} + 3\mathcal{H}\left(\rho + \press \right) + \boldsymbol{\nabla} \cdot \left(\rho + \press \right) \vel = 0\label{eq:cont_rf}\\
&\der{\vel}{\tau} + \mathcal{H}\vel + \left( \vel \cdot  \boldsymbol{\nabla} \right)  \vel + \boldsymbol{\nabla} \phi = -  \frac{c_s^2\boldsymbol{\nabla} \rho}{\rho + \press} \label{eq:eul_rf}\\
&\frac{p}{c^2}= w \bar{\rho} + \frac{c_s^2}{c^2} \delta\rho \,.
\label{eq:eos_rf}
\end{align}
Notice that in the limit $c_s\to 0$, the Euler equation is equal to that for pressureless matter. This shows that, assuming adiabatic initial conditions, matter and dark energy co-move on large scales. 
By writing $\rho = \bar\rho (1+\delta_{\rm de})$, one can relate $\delta_{\rm de}$ to the quantity $\delta_A$ defined in \cite{2024JCAP...03..032D} via
$\delta_{\rm de} = (\bar\rho_{\rm m,0}/\bar\rho_{\rm de,0}) a^{3w} [\delta_A - \delta_{\rm m}]$.
It is straightforward to show that, in the $c_s\to 0$ limit, \refeq{cont_rf} agrees with the nonlinear continuity equation for $\delta_A$ derived there in the same limit.

To ease the implementation, following~\citet{Blot:2022mnt}, we define a new variable
\be
\Pi=\frac{\rho+\press}{\bar{\rho}}=1+w+\left(1+\cssqnorm\right)\,\delta,
\ee
which allows us to write the equations in quasi-conservative form
\begin{align}
&\der{\Pi}{\tau} + \left( 1+\frac{c_s^2}{c^2} \right) \boldsymbol{\nabla} \cdot \left(\Pi \vel \right) =  3 \mathcal{H} \left(w - \frac{c_s^2}{c^2}\right)\left(\Pi-1-w \right)\,, \\
&\der{\Pi \vel}{\tau} + \mathcal{H}\,\Pi\vel + \boldsymbol{\nabla} \cdot \left( \Pi \vel \otimes \vel \right) + \frac{c_s^2\, \boldsymbol{\nabla} \Pi}{\left(1+\frac{c_s^2}{c^2}\right)} =3\mathcal{H}w\,(\Pi-1-w)\vel- \Pi\, \boldsymbol{\nabla}\phi\,,\label{eq:eul_qc}
\end{align}
where we have dropped higher-order source terms of the Euler equation that are suppressed by $c_s^2/c^2$. 

Finally, we define the super-conformal units as:
\begin{eqnarray}
\frac{d}{d\tau}=\frac{1}{a\;t_*}\frac{d}{d\tilde{t}} \\
\frac{d}{dx}=\frac{1}{r_*}\frac{d}{d\tilde{r}} \\
v=\frac{v_*}{a}\tilde{v} \\
\phi = \frac{\phi_*}{a^2}\tilde{\phi}
\end{eqnarray}
where
\begin{eqnarray}
r_*=v_*t_* \\
\phi_*=v_*^2
\end{eqnarray}
so the equations that we implement in \nefertiti are
\ba
&\der{\Pi}{\tilde{t}} + \left( 1+\frac{c_s^2}{c^2} \right) \tilde{\boldsymbol{\nabla}} \cdot \left(\Pi\tilde{\vel} \right) = 3 \tilde{H} \left(w - \frac{c_s^2}{c^2}\right)\left(\Pi-1-w \right)\,,\label{eq:cont_sc}\\
&\der{\Pi \tilde{\vel}}{\tilde{t}} + \tilde{\boldsymbol{\nabla}} \cdot \left( \Pi \tilde{\vel} \otimes \tilde{\vel} \right) + \frac{\tilde{c_s}^2 \tilde{\boldsymbol{\nabla}} \Pi}{\left(1+\frac{c_s^2}{c^2}\right)} = 3 \tilde{H}w\left(\Pi-1-w \right)\tilde{\vel}- \Pi \tilde{\boldsymbol{\nabla}}\tilde{\phi}\,,
\label{eq:eul_sc}\ea
where tildas indicate super-conformal units. On the right-hand side of \refeq{eul_sc} a $\tilde{\vel}$ appears, which is not one of our independent variables in the conservative formulation. We compute it as $\Pi \tilde{\vel}/\Pi$.

\section{Integration of the source terms}\label{sec:sources}
Equations \refeqs{cont_sc}{eul_sc} contain cosmological source terms that depend on the Hubble parameter, which are not present in the usual baryonic fluid equations. To integrate these terms numerically, we follow the same approach that is used in \texttt{Ramses} and employ source splitting, where the full solution is achieved by solving separately the ordinary differential equations (ODEs)
\ba
&\frac{\mathrm{d}\Pi}{\mathrm{d}\tilde{t}}=3 \tilde{H} \left(w - \frac{c_s^2}{c^2}\right)\left(\Pi-1-w \right)\,,\label{eq:cont_ode}\\
&\frac{\mathrm{d}\Pi \tilde{\vel}}{\mathrm{d}\tilde{t}}=3 \tilde{H}w\left(\Pi-1-w \right)\tilde{\vel}- \Pi \tilde{\boldsymbol{\nabla}}\tilde{\phi}\label{eq:eul_ode}\,,
\ea
and the advection equations
\ba
&\der{\Pi}{\tilde{t}} + \left( 1+\frac{c_s^2}{c^2} \right) \tilde{\boldsymbol{\nabla}} \cdot \left(\Pi\tilde{\vel} \right) = 0\,,\label{eq:cont_adv}\\
&\der{\Pi \tilde{\vel}}{\tilde{t}} + \tilde{\boldsymbol{\nabla}} \cdot \left( \Pi \tilde{\vel} \otimes \tilde{\vel} \right) + \frac{\tilde{c_s}^2 \tilde{\boldsymbol{\nabla}} \Pi}{\left(1+\frac{c_s^2}{c^2}\right)} = 0\,.\label{eq:eul_adv}\ea
Second order accuracy in time is achieved by using Strang splitting, which consists in advancing the sources by $\Delta \tilde{t}/2$, where $\Delta \tilde{t}$ is the time-step, then advancing the advection by a full time-step $\Delta \tilde{t}$ and finally updating the sources by another half time-step $\Delta \tilde{t}/2$.\\

To integrate the cosmological source terms, we further split \refeq{eul_ode} into the cosmological and gravitational parts, where the integration of the gravitational source term is handled numerically with a predictor-corrector scheme, while the cosmological source term is integrated analytically. The analytical solution of the ODEs
\ba
&\frac{\mathrm{d}\Pi}{\mathrm{d}\tilde{t}}=3 \tilde{H} \left(w - \frac{c_s^2}{c^2}\right)\left(\Pi-1-w \right)\,,\\
&\frac{\mathrm{d}\Pi \tilde{\vel}}{\mathrm{d}\tilde{t}}=3 \tilde{H}w\left(\Pi-1-w \right)\tilde{\vel}
\ea
can easily be found by changing the time variable to the scale factor $a$ and using
\be
\tilde{H}=\frac{1}{a}\frac{\mathrm{d}a}{\mathrm{d}\tilde{t}}\,,
\ee
so that 
\ba
&\frac{\mathrm{d}\Pi}{\Pi-1-w}=3\left(w-\cssqnorm\right)\frac{\mathrm{d}a}{a}\,,\label{eq:cont_source}\\
&\frac{\mathrm{d}\Pi\tilde{\vel}}{\Pi\tilde{\vel}}=3w\left(\frac{\Pi-1-w}{\Pi}\right)\frac{\,\mathrm{d}a}{a}\,.\label{eq:mom_source}
\ea
Integrating on both sides of \refeq{cont_source} we get
\be
\Pi(a)=1+w+a^{3\left(w-\cssqnorm\right)}K\,,
\ee
where $K$ is an integration constants. We can determine this constant by using the initial conditions, so that the final expression for the solutions is
\be
\Pi(a)=1+w+\left(\frac{a}{a_{\rm ini}}\right)^{3\left(w-\cssqnorm\right)}\Pi_{\rm ini}\,,
\ee

To find a solution for \refeq{mom_source} we need an expression for $\Pi(a)$. We approximate the time evolution of $\Pi$ within the timestep as the linear solution in Einstein-de Sitter cosmologies, which is directly proportional to that of matter, i.e.
\be
\delta = \delta_{\rm ini} \frac{a}{a_{\rm ini}}.
\ee
With this, we obtain:
\be
\frac{d\ln \Pi\vel}{d\ln a} = 3 w \frac{\left(1+\cssqnorm\right)\frac{\delta_{\rm ini}}{a_{\rm ini}}a}{1+w+\left(1+\cssqnorm\right)\frac{\delta_{\rm ini}}{a_{\rm ini}} a}\,,
\ee
which can be integrated to yield
\be
\ln \Pi\vel - \ln \Pi\vel_{\rm ini} = 3 w \ln\left(1+w+\left(1+\cssqnorm\right)\delta_{\rm ini} a/a_{\rm ini}\right)\Big|^a_{a_{\rm ini}}\,,
\ee
or
\be
\Pi\vel =\Pi\vel_{\rm ini} \left(\frac{1+w+\left(1+\cssqnorm\right)\delta_{\rm ini} a/a_{\rm ini}}{1+w+\left(1+\cssqnorm\right)\delta_{\rm ini}}\right)^{3w}=\Pi\vel_{\rm ini} \left(\frac{1+w+\left(1+\cssqnorm\right)\delta_{\rm ini} a/a_{\rm ini}}{\Pi_{\rm ini}}\right)^{3w}\,.
\ee
Notice that this solution might diverge when $\Pi_{\rm ini}\rightarrow 0$. Since this does not occur in the simulations presented here, we leave the treatment of this regime for future work.

\end{document}